\documentclass[nofootinbib,pra,superscriptaddress,a4paper,twocolumn]{revtex4-1}
\usepackage{geometry}
\usepackage[english]{babel}
\usepackage[latin1]{inputenc}
\usepackage{hyperref}
\usepackage{amsmath, amssymb, amsfonts, mathrsfs}
\usepackage{graphicx}
\usepackage{multirow}
\usepackage{color, colortbl}
\usepackage{varwidth}
\definecolor{mycolor}{rgb}{0.6, 0.8, 0.2}
\usepackage{capt-of}
\usepackage{diagbox}
\newcommand{\ket}[1]{|#1\rangle}
\newcommand{\bra}[1]{ \langle #1 \,  |}

\begin{document}

\title{Quantum Experiments and Hypergraphs: Multi-Photon Sources for Quantum Interference, Quantum Computation and Quantum Entanglement}

\author{Xuemei Gu}
\email{xmgu@smail.nju.edu.cn}
\affiliation{State Key Laboratory for Novel Software Technology, Nanjing University, 163 Xianlin Avenue, Qixia District, 210023, Nanjing City, China.}

\author{Lijun Chen}
\email{chenlj@nju.edu.cn}
\affiliation{State Key Laboratory for Novel Software Technology, Nanjing University, 163 Xianlin Avenue, Qixia District, 210023, Nanjing City, China.}

\author{Mario Krenn}
\email{mario.krenn@univie.ac.at}
\affiliation{Department of Chemistry \& Computer Science, University of Toronto, Canada \& Vector Institute for Artificial Intelligence, Toronto, Canada.}

\begin{abstract} 
We introduce the concept of hypergraphs to describe quantum optical experiments with probabilistic multi-photon sources. Every hyperedge represents a correlated photon source, and every vertex stands for an optical output path. Such general graph description provides new insights for producing complex high-dimensional multi-photon quantum entangled states, which go beyond limitations imposed by pair creation via spontaneous parametric down-conversion. Furthermore, properties of hypergraphs can be investigated experimentally. For example, the \textit{NP-Complete} problem of deciding whether a hypergraph has a perfect matchin can be answered by experimentally detecting multi-photon events in quantum experiments. By introducing complex weights in hypergraphs, we show a general many-particle quantum interference and manipulating entanglement in a pictorial way. Our work paves the path for the development of multi-photon high-dimensional state generation and might inspire new applications of quantum computations using hypergraph mappings.
\end{abstract}
\date{\today}
\maketitle

\section{Introduction}

Graph-theoretical concepts are widely used in multidisciplinary research involving physics, chemistry, neuroscience and computer sciences, among others. Considerable progress has taken place in recent years in the direction of applying graph theory in quantum physics. Their connections have been carried out explicitly leading to many interesting and complementary works. A well-known example is the so-called graph states, the structure of which can be described in a concise and fruitful way by mathematical graphs \cite{hein2004multiparty}. These states form a universal resource for quantum computing based on measurements \cite{raussendorf2001one, raussendorf2003measurement} and later has been generalized to continuous-variable quantum computation \cite{menicucci2006universal}, using an interesting connection between gaussian states and graphs \cite{menicucci2011graphical}. Graphs have also been used to study collective phases of quantum systems \cite{shchesnovich2018collective}, characterize quantum correlations \cite{cabello2014graph} and investigate quantum random networks \cite{perseguers2009quantum, cuquet2009entanglement}. 

A graph theoretical approach to quantum mechanics would also help to amalgamate visualization offered by graphs with the well developed mathematical machinery of graph theory. For example, a very powerful pictorial tool for making calculations in quantum mechanical theory is Feynman diagrams, which is indispensable in the context of quantum electrodynamics \cite{feynman1949new}. Moreover, quantum processes such as teleportation,  logic-gate teleportation, entanglement swapping, etc. can be captured at a more abstract level \cite{coecke2017picturing, abramsky2004categorical}. A different graphical representation  has been developed to describe quantum states and local unitaries \cite{dutta2016graph}. Also, directed graphs have recently been investigated in order to simplify certain calculations in quantum optics, by representing creation and annihilation operators in a visual way \cite{ataman2018graphical}. 

Recently, an entirely different method of connecting graphs in quantum physics has been introduced, by showing that Graphs can capture essential elements of quantum optical experiments \cite{krenn2017quantum, gu2019quantum, gu2019quantum3}. The graph-experiment connection exploits the fact that the most common sources of photonic entanglement are spontaneous parametric down-conversion (SPDC) \cite{hardy1992source}, which is a nonlinear process that probabilistic creates photon pairs. Those photon pairs are then interpreted as two vertices connected by an edge. This simple idea has been exploited in \cite{krenn2017quantum, gu2019quantum, gu2019quantum3} to understand better the generalization of quantum states, quantum information protocols and for gaining new insights towards quantum computation.

Can such graph-experiment concept be extended and applied in the quantum experiments with other probabilistic photon sources that are not restricted to pairs of photons such as SPDC? The answer is positive and this motivates us to provide a general description for quantum experiments using different probabilistic photon sources, for example experimentally single-photon sources in form of the attenuated lasers or sources that can produce $n>2$ photon tuples. For convenience, we call all these probabilistic sources as  $n$-photon sources, which produces $n$ correlated photons for the rest of the paper. Therefore, a single-photon source in the form of attenuated lasers is just a 1-photon source, and the SPDC-based nonlinear crystal is a 2-photon source.

\begin{figure}[!t]
	\includegraphics[width=0.44\textwidth]{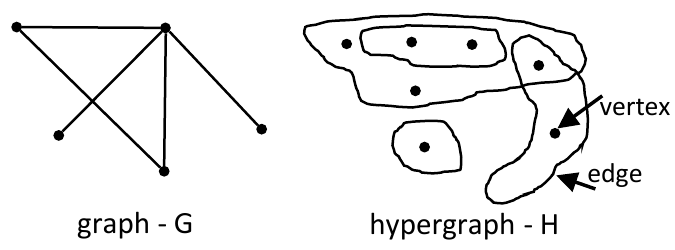}
	\caption{An example of graphs and hypergraphs. In a graph, an edge connects only a pair of vertices, but an edge of a hypergraph -- known as hyperedge -- can connect an arbitrary number of vertices.}
	\label{fig:HypergraphConcept}
\end{figure}

\begin{table}[!t]
	\centering
	\caption{The analogies between hypergraphs and quantum experiments.}
	\begin{tabular}{p{3.9cm}|p{3.9cm}}
		\hline 
		\textbf{Hypergraphs}&\textbf{Quantum Experiments}\\ \hline
		undirected hypergraphs& quantum optical setup with $n$-photon sources\\ \hline
		vertex&optical output path\\ \hline
		hyperedge& $n$-photon sources\\ \hline
		colors in the boundary of the hyperedge &photon's mode numbers \\ \hline
		colors in the region of the hyperedge & phases between photons \\ \hline
		transparency in the region of the hyperedge & amplitudes of photons  \\ \hline
		perfect matchings & $N$-fold coincidences \\ \hline
	\end{tabular}
	\label{tab:hypersteup}
\end{table}

Hypergraphs are generalizations of graphs \cite{berge1973graphs, bretto2013hypergraph}, which have caused a significant interest in applications to real-world problems. Here we show that hypergraphs are a suitable mathematical model and effective tool for quantum experiments. Our contributions are: 1) We introduce a mapping between hypergraphs and quantum experiments, and properties of hypergraphs capture properties of the experiments (Section II). 2) We show how this mapping can be used to design quantum experiments in an abstract and systematic way (Section III). 3) Thereby, we find entirely concepts of new setups for the generation of complex quantum states, using a combination of 1- and 2-photon sources. Those setups can be performed experimentally with standard quantum optics technology, and use less resources than state-of-the-art techniques (Section IIIA). 4) With the mapping, we also find new experimental configurations that overcome limitations from linear-optics experiments using SPDC crystals. It allows to produce much higher-dimensionally entangled Greenberger-Horne-Zeilinger (GHZ) states than possible with SPDC (Section IIIB). 5) The abstract structure of hypergraphs for describing experiments let us identify mathematical challenges in hypergraphs and translate them to experimental setups. In particular, we show that hypergraph-generalizations of BosonSampling (which is in different complexity classes as the standard approaches) can simply be designed using the new hypergraph mapping (Section IV). 6) We show that hypergraphs can be used to understand interference structures in complex experiments intuitively with pictures. Thereby, we reinterpret the results of a previously introduced quantum interference experiment (Section V).

\section{Quantum Experiments and Hypergraphs}

Here we briefly review some basic notation and terminology of hypergraphs that will be useful later. Formally speaking, a hypergraph $H$ is a pair $H=\{V,E\}$, where $V=\{v_{i}|i=1,2,...,x\}$ is the set of $x$ vertices (or nodes) and $E=\{e_{i}|i=1,2,...,y\}$ is the set of $y$ hyperedges. The number of vertices that a hyperedge $e$ contained (or the degree of a hyperedge) is denoted as $d(e)$ and the number of hyperedges that a vertex $v$ involved (or the degree of a vertex) is denoted as $d(v)$. A hypergraph is called $k$-uniform, if every hyperedge contains exactly $k$ vertices, i.e., $d(e)=k,\forall e \in E$. Clearly, an ordinary graph is a special case of hypergraphs where $d(e)=2$, namely a \textit{2-uniform hypergraph}. An example of graphs and hypergraphs is given in Fig. \ref{fig:HypergraphConcept}.

\begin{figure}[!t]
	\includegraphics[width=0.5\textwidth]{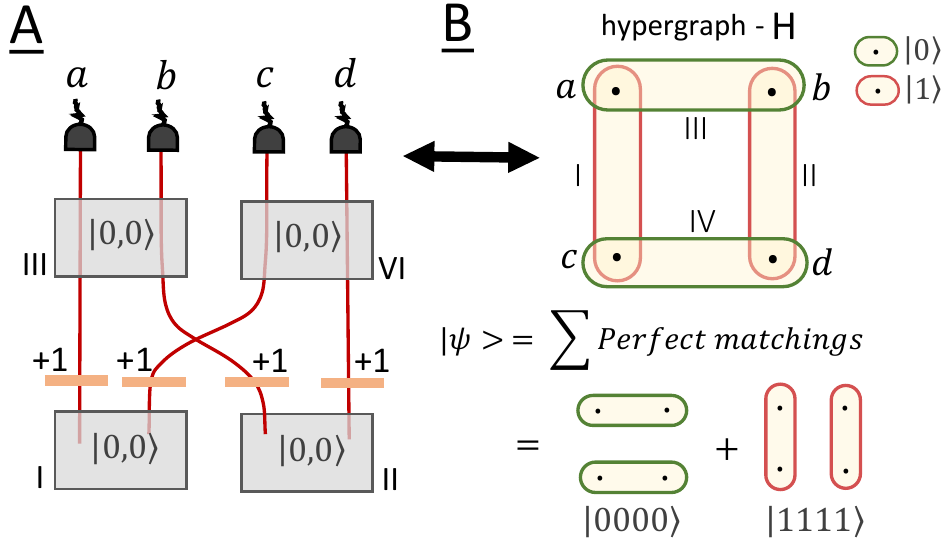}
	\caption{Hypergraph-Experiment link for producing 2-dimensional 4-particle GHZ states. \textbf{A}: Four nonlinear crystals (gray squares) are pumped coherently, and the pump laser can be set such that two photon pairs are created with reasonable probabilities. A photon's mode numbers can be changed by inserting variable mode shifters (yellow squares) in the photon's path. The final quantum state is obtained when four detectors click simultaneously. \textbf{B}: Its hypergraph $H$ with four vertices (one for a photon's output path) and four hyperedges (one for a crystal). The color of the boundary of a hyperedge stands for the photons' mode numbers, such as red is $\ket{1}$ and green $\ket{0}$. The color (for instances light yellow) in the region of the hypergraph depicts global phases. A four-fold coincidence event is described as a perfect matching of the hypergraph, which is \textit{a collection of edges covering all vertices only once}. Thus the post-selection quantum state is given as $\ket{\psi}=\frac{1}{\sqrt{2}}(\ket{0000}+\ket{1111})$.}
	\label{fig:HypergraphLink}
\end{figure}

Now we start with a simple quantum experiment to illustrate how we update the graph-experiment link to hypergraphs in Fig. \ref{fig:HypergraphLink}. There all nonlinear crystals are pumped coherently. As the SPDC process is entirely probabilistic, it means the probability of obtaining two pairs from one crystal or one pair from two crystals is the same. Such situations and even multiple pair emissions from SPDC are unavoidable. However, one can adjust the pump power such that these cases are in a sufficiently low probability which can be safely neglected. Therefore, we can adjust the laser power such that only two photon pairs are created with reasonable probabilities \cite{krenn2017entanglement}. Every photon path represents a vertex and every nonlinear crystal corresponds to an edge \cite{krenn2017quantum, gu2019quantum3}. An $N$-fold coincidences event ($N$ detectors click simultaneously) is described as a perfect matching -- a subset of edges that visit all vertices exactly once. Thus the final quantum state under the condition of $N$-fold coincidences is given as the coherent superposition of perfect matchings in the graph. As a regular graph is just a \textit{2-uniform hypergraph}, we can reinterpret all the graph representations into hypergraphs. Thus a nonlinear crystal is a hyperedge with $d(e)=2$ and an optical path is a vertex in a hyperedge (see the correspondence listed in Table. \ref{tab:hypersteup}). The hypergraph interpretation of the experiment is shown in Fig. \ref{fig:HypergraphLink}B.

There are two perfect matchings in the hypergraph -- \textit{a collection of edges covering all vertices only once}, indicating that 4-fold coincidences happen when crystals (I$\&$II) or (III$\&$VI) fire together. Thus the resulting quantum state is $\ket{\psi}=\frac{1}{\sqrt{2}}(\ket{0000}+\ket{1111})$, where numbers 0 and 1 denote as photons' mode numbers that correspond to the polarization of photons, or high-dimensional degrees of freedoms \cite{erhard2019advances} such as the orbital angular momentum (OAM) \cite{allen1992orbital, rubinsztein2016roadmap, krenn2017orbital}, time bin \cite{franson1989bell, versteegh2015single} or frequency \cite{olislager2010frequency}.

\section{State Generation}

The original graph description is only applicable to model pair correlations (2-photon source case, $n=2$) , thus it cannot be used to describe other photon sources such as 1-photon sources  where $n=1$. However, using our hypergraph-experiment technique, we can describe arbitrary probabilistic source, and their combinations.

The power of our technique is that it allows us to design new quantum experiments, using clear mathematical structures. If one wants to find an experimental setup for a specific quantum entangled state, one can now rephrase the problem in terms of (hyper)graph theory. A specific state corresponds to a hypergraph with specific perfect matchings. Identifying perfect matchings is a task that is oftentimes easier than finding the experimental setup itself, for example, because it can be formalized as a simple mathematical question that can be solved by standard mathematical software. A solution to the mathematical question can then be directly translated to the solution of the experimental setup. Interestingly, if a solution cannot be found mathematically, there cannot be an experimental setup that produces the target state with the provided resources.

We show this design principle on several simple examples, which require hypergraphs. However, the formalism could apply more broadly in the same way.

Here we investigate state generations with $n$-photon sources where $n$ is not necessary $n=2$. First, we study the case $n=1$, which has been experimentally implemented several times in the form of attenuated lasers\cite{eisaman2011invited}. We find efficient and compact setups for high-dimensional multiphotonic state generations. Then we continue with the case $n>2$ and find that one can overcome limitations for the dimensionality of multiphotonic states.

\begin{figure}[!t]
	\includegraphics[width=0.5\textwidth]{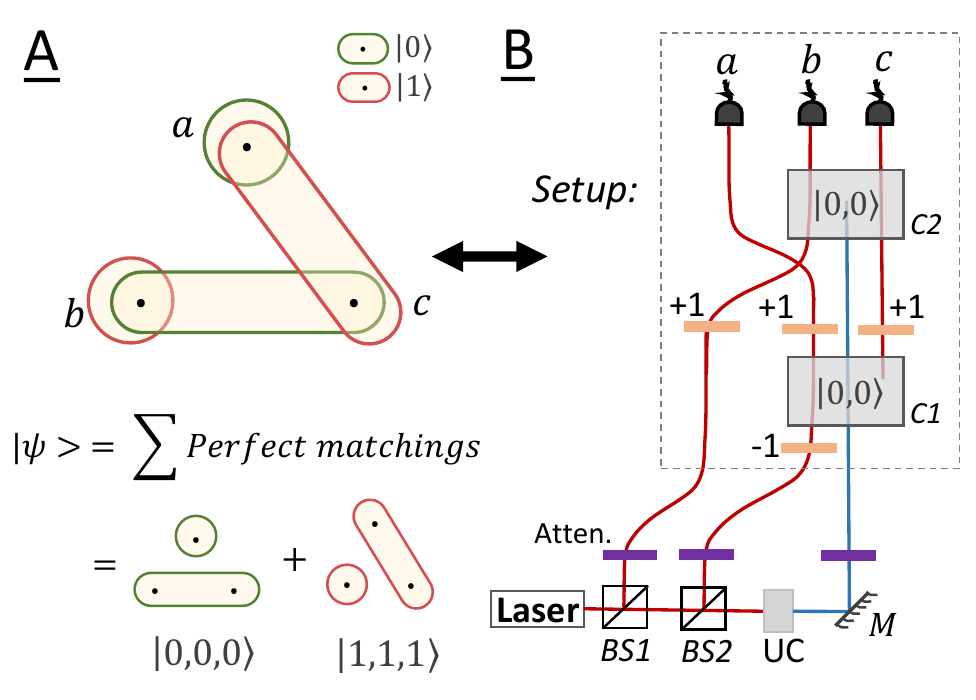}
	\caption{Hypergraph-Experiment link for producing 2-dimensional 3-particle GHZ states. \textbf{A}: A hypergraph with 4 hyperedges and 3 vertices. There are two perfect matchings -- \textit{a subset of hyperedges visiting all vertices exactly once}, and each stands for one term in the expected quantum states. Hyperedges of $d(e)=2$ and $d(e)=1$ represent nonlinear crystals and laser-generated single photons respectively. Two perfect matchings lead to 3-fold coincidences, which leads to the final post-selected state $\ket{\psi}=\frac{1}{\sqrt{2}}(\ket{000}+\ket{111})$. \textbf{B}: The corresponding setup (dashed part). An infrared laser (red lined) after two BSs is used to achieve up-conversion (UC) with high efficiency \cite{zhao2004experimental}. The UC produced light beam (blue lined) are subsequently exploited for photon pair creation via SPDC. With the hypergraph-experiment link, one can make output paths identical using \textit{path identity} \cite{krenn2017entanglement}, which is fully experimentally feasible as results shown in \cite{kysela2019experimental}. The attenuators (purple lined) are used for tuning the power in order to change the photon creation probability. }  
	\label{fig:LaserGHZ2d}
\end{figure}

\begin{figure*}[!t]
	\includegraphics[width=0.98\textwidth]{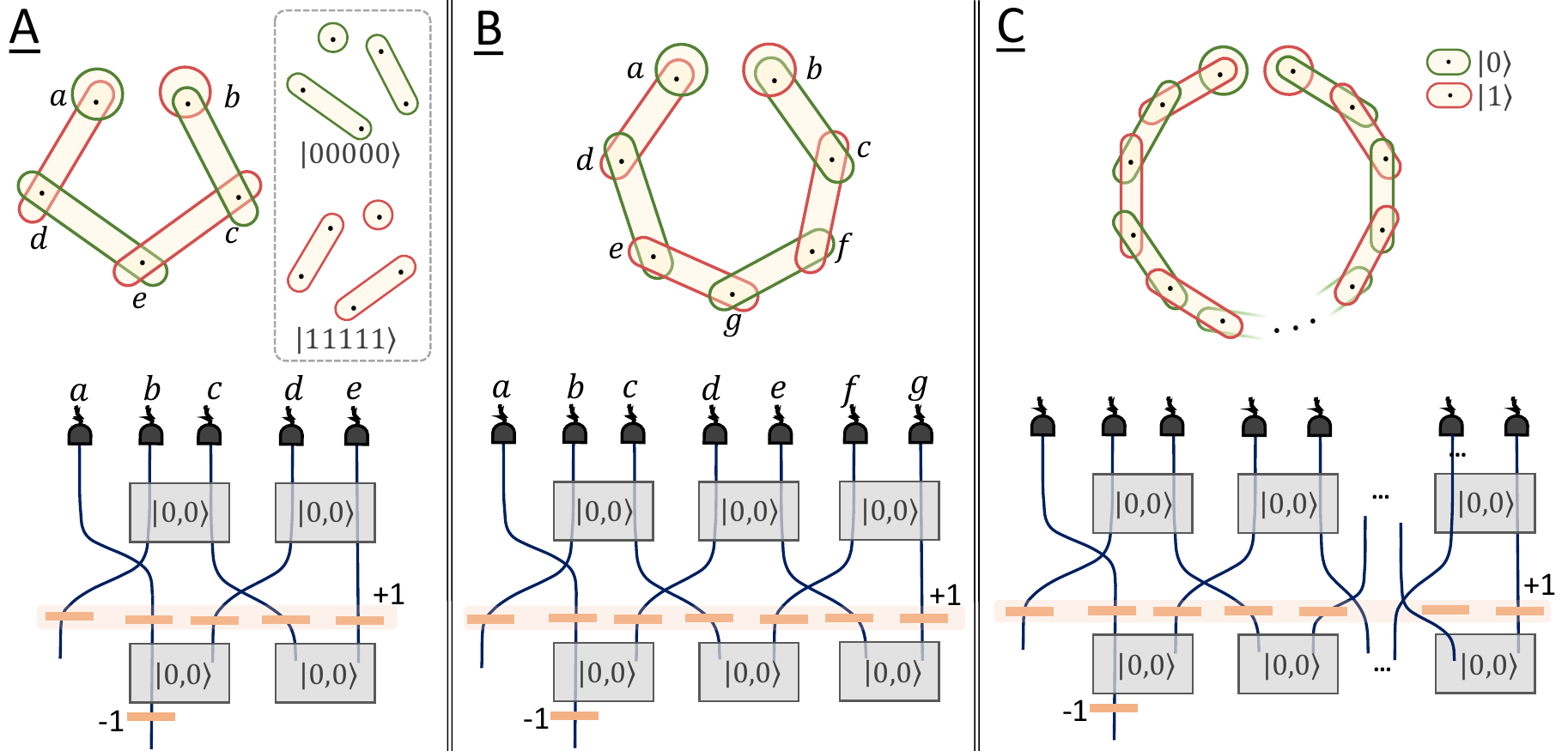}
	\caption{General hypergraphs and setups for producing 2-dimensional (odd) $m$-particle GHZ states using 1-photon and 2-photon sources. In Fig.\ref{fig:LaserGHZ2d}A, we show a hypergraph for $m=3$. One can arbitrarily extend that hypergraph without introducing new perfect matchings by adding more hyperedges $d(e)=2$, which means inserting several 2-photon sources. \textbf{A}-\textbf{C} show general hypergraphs and setups for creating 2-dimensional 5-, 7-, $m$-particle GHZ states.}  
	\label{fig:oddGHZ}
\end{figure*}

\begin{figure}[!t]
	\includegraphics[width=0.5\textwidth]{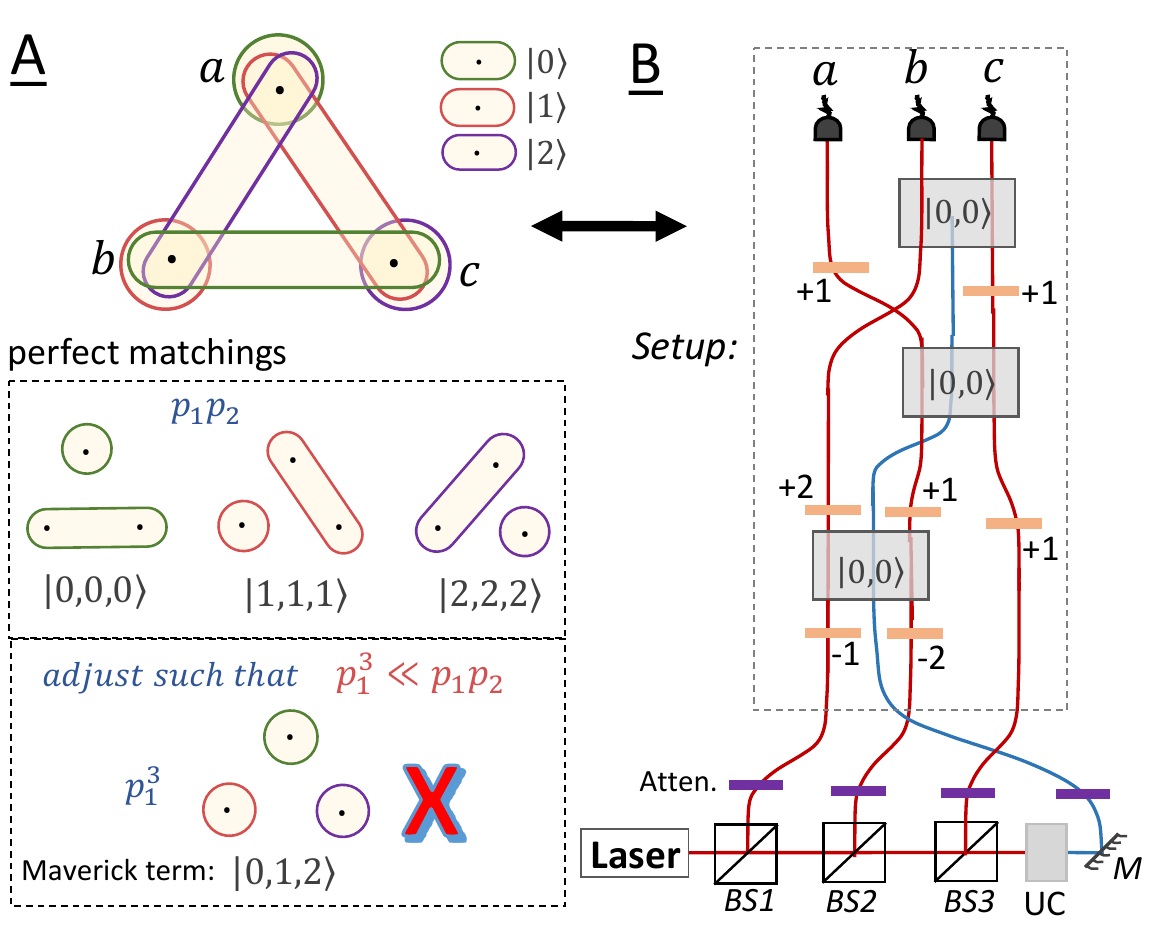}
	\caption{Hypergraph-Experiment link for producing 3-dimensional 3-particle GHZ states. \textbf{A}: Analogous to Fig.\ref{fig:LaserGHZ2d}A, we add one more perfect matching which relates to quantum state term $\ket{222}$. However, we find that there will be one additional perfect matching for the term $\ket{012}$ (\textit{Maverick term}). The probabilities of existence for hyperedges $d(e)=1$ and $d(e)=2$ are $p_{1}$ and $p_{2}$, which can be experimentally adjusted by attenuators. Thus one can make the Maverick term in a sufficiently low probability or negligible if $p_{1}^{3} \ll p_{1}p_{2}$. \textbf{B}: The corresponding experimental implementation. This setup is a specially resource providing us a efficient and feasible technique to produce high-dimensional GHZ states.}  
	\label{fig:LaserGHZ3d}
\end{figure}

\begin{figure*}[!t]
	\includegraphics[width=\textwidth]{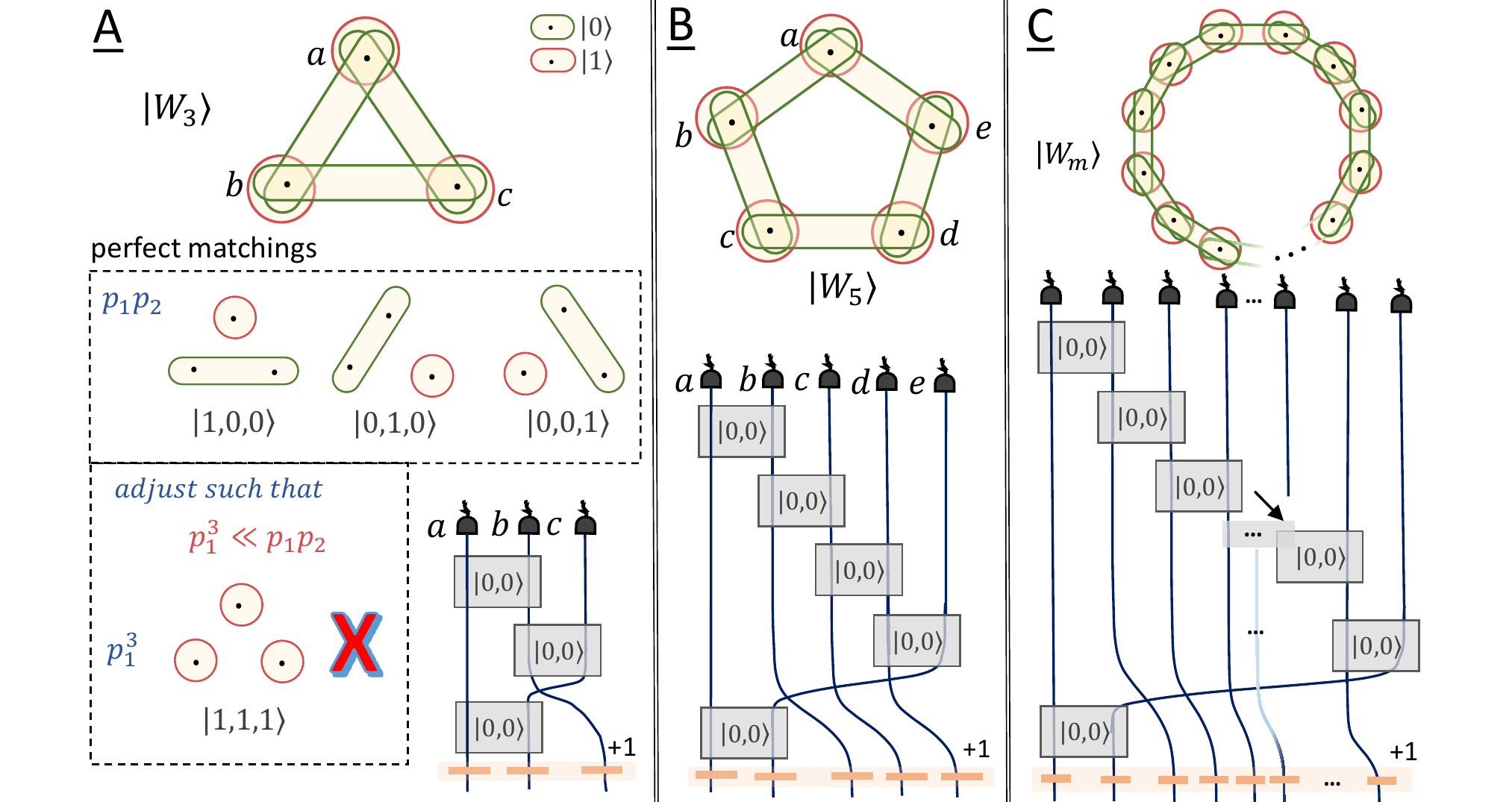}
	\caption{General hypergraphs and setups for producing (odd) $m$-particle W states. \textbf{A:} Analogous to Fig.\ref{fig:LaserGHZ3d}A, one can adjust $p_{1}$ and $p_{2}$ to reduce the probability of existence of perfect matchings. We depicted the boundaries of all hyperedges $d(e)=1$ and $d(e)=2$ in red and green, respectively. Thus every perfect matching contains only one red bounded hyperedge $d(e)=1$, which means every term in the quantum state has exactly one excitation. The coherent superposition of perfect matchings describes a 3-particle W state. The related experimental setup is shown below the hypergraph. \textbf{B}-\textbf{C} describe general hypergraphs and setups for 5-, $m$-particle W state under the condition of $p_{1}^2 \ll p_{2}$.}  
	\label{fig:oddW}
\end{figure*}

\subsection{$n$-photon sources: $n<3$}
\subsubsection{GHZ states}
GHZ states are the most prominent example for non-classical correlations between more than two involved parties, and have led to new understanding of the fundamental properties of quantum physics \cite{greenberger1989going, greenberger1990bell}. Such multipartite quantum entanglement are denoted as
\begin{equation}
\ket{GHZ_{m}^{d}}=\frac{1}{\sqrt{d}}\sum_{i=0}^{d-1} \ket{i}^{\bigotimes n}
\end{equation}
where $m$ is the number of particles and $d$ is the dimension for every particle.

For a 2-dimensional 3-particle GHZ state
\begin{equation}
\ket{GHZ_{3}^{2}}=\frac{1}{\sqrt{2}}(\ket{000}+\ket{111}),
\end{equation}
we show how to construct a hypergraph which describes the target state $\ket{GHZ_{3}^{2}}$. Three particles indicate that there are three vertices in the hypergraph. There are two terms in the quantum state, meaning that there will be two perfect matchings in the hypergraph. One stands for term $\ket{000}$, and the other represents term $\ket{111}$.  In Fig. \ref{fig:LaserGHZ2d}A, we show such a hypergraph with two perfect matchings consisted of hyperedges $d(e)=1$ and $d(e)=2$.  These two perfect matchings contribute to 3-fold coincidences cases in the experiment, and their coherent superposition leads to the final entangled state, which is our expected 2-dimensional 3-particle GHZ state. 

Equipped with the hypergraph-experiment connection from Table. \ref{tab:hypersteup}, we can then transfer the hypergraph into its corresponding experimental implementation. Hypergraphs of $d(e)=1$ and $d(e)=2$ indicates that we use $n$-photon sources ($n=1,2$) in the setup, and the vertices represent photon output paths, see Fig. \ref{fig:LaserGHZ2d}B. There two beam splitters (BS) are properly used to separate an infrared pulse laser beam into three paths, where the single photon is approximated by an attenuated laser beam. The mostly transmitted laser beam undergoes up-conversion process (UC) for producing blue light with high efficiency \cite{zhao2004experimental}, which are used to coherently pump two nonlinear crystals to generate correlated photon pairs with reasonable probabilities. Then the laser-generated single photon and photon pairs output paths are aligned identically (namely \textit{path identity} \cite{krenn2017entanglement}) such that one cannot determine the which-source information when observing the detectors. 

One can experimentally adjust the power of the infrared laser and exploit attenuators to change the photon creation probability $p_{i}, (i=1,2)$, where subscript $i$ stands for the used photon source. In the case of $i=1$, the used photon source is 1-photon source. In the case of $i=2$, that corresponds to a 2-photon source. Instead of using heralded single photons via SPDC to produce 3-particle entanglement with probability $p_{2}^{2}$, we can obtain the expected quantum state with probability $p_{1}p_{2}$.

Now we generalize the hypergraph for producing 2-dimensional $m$-particle GHZ states using $n$-photon sources. For even $m$, we can use $n=2$ only \cite{gu2019quantum3} while for odd $m$, we show that $n=1$ together with several $n=2$ works in Fig. \ref{fig:oddGHZ}. In addition to multiple GHZ states, we can also produce 3-dimensional GHZ states, shown in Fig. \ref{fig:LaserGHZ3d}. This is an alternative approach to the recent experimental implementation based on linear optics \cite{erhard2018experimental}. It is a very compact, reasonable experimental scheme for a high-dimensional GHZ state that can be realized in laboratories. 
 
\subsubsection{W states}
We further illustrate our approach on W states, an important multiphoton entanglement class that is highly persistent against photon loss \cite{zeilinger1992higher, bourennane2004experimental}. W states are one specially case of Dicke states introduced by Robert H. Dicke \cite{dicke1954coherence}, which cannot be transformed into GHZ states with
local operation and classical communication (LOCC) \cite{dur2000three} and  defined as
\begin{equation}
\ket{W_{m}}=\frac{1}{\sqrt{m}}\hat{S}(\ket{0}^{\bigotimes(n-1)}\ket{1})
\end{equation}
where $m$ stands for the number of particles and $\hat{S}$ is the symmetrical operator that gives summation over all distinct permutations of the $m$ particles.

Let us consider the simplest case with a 3-particle W state
\begin{equation}
\ket{W_{3}}=\frac{1}{\sqrt{3}}(\ket{001}+\ket{010}+\ket{100}),
\end{equation}
and show how one can exploit the hypergraph-experiment connection. There are three terms in the quantum state, which correspond to three perfect matchings in the hypergraph. 

Analogous to Fig. \ref{fig:LaserGHZ3d}A, we construct a hypergraph for the W state under the condition $p_{1}^{2} \ll p_{2}$ in Fig. \ref{fig:oddW}A. Different colors of the boundaries stand for different mode numbers of photons, for example red means mode number $1$. We observed that every perfect matching in the hypergraph involves only one red bounded hyperedge $d(e)=1$. That means every term in the quantum state contains exactly one excitation and their coherent superposition describes a 3-particle W state. Then we generalize hypergraphs for $m$-photon W states ($m$ is odd number) by \textit{path identity} in theory under the condition of $p_{1}^{2} \ll p_{2}$ (on-chip scheme has been studied in \cite{fengchipone}, there the authors use two 2-photon sources where one of the 4 photons heralded W states.). There each vertex is joined by a red bounded hyperedge and every two vertices are connected by a green bounded hyperedge, shown in Fig. \ref{fig:oddW}B and C. It ensures that every perfect matching in the hypergraph gives exactly one excitation in every term of the quantum state.

\subsubsection{Schmidt-Rank Vector states}
When quantum entangled states go to higher dimensions, interesting properties \cite{lawrence2014rotational} and non-classical correlations \cite{huber2013structure, huber2013entropy, cadney2014inequalities} will show up. Those new structures of multipartite high-dimensional entanglement are characterized by the Schmidt-Rank Vector (SRV) and give rise to new phenomena that only exist if both the number of particles and the number of dimensions are larger than two \cite{huber2013structure, huber2013entropy, cadney2014inequalities}.
 
The SRV represents the rank of the reduced density matrices of each particle. In the three-particle pure state case (which is the case we consider and the three parties are $a$, $b$ and $c$), the rank of the reduced density matrices
\begin{align}
A=rank(Tr_{a}(\ket{\psi}\bra{\psi}))\nonumber\\
B=rank(Tr_{b}(\ket{\psi}\bra{\psi}))\\
C=rank(Tr_{c}(\ket{\psi}\bra{\psi}))\nonumber
\label{eq:SRVrank}
\end{align}

together form the SRV $d_{\psi}=(A, B, C)$, where $A\geq B\geq C$. The values $A$, $B$ and $C$ stand for the dimensionality of entanglement (Schmidt-Rank) between every particle with the other two parties (for example $a$-$bc$, $b$-$ac$ and $c$-$ab$ correspond to the entanglement between $a$ and $bc$, $b$ and $ac$, $c$ and $ab$). The classification with different SRVs provides an interesting insight that one can transform quantum states from higher classes to lower classes with LOCC, and not vice versa, which means the dimensionality $i$ ($i=A, B, C$) cannot be increased with LOCC.

\begin{figure}[!t]
	\includegraphics[width=0.5\textwidth]{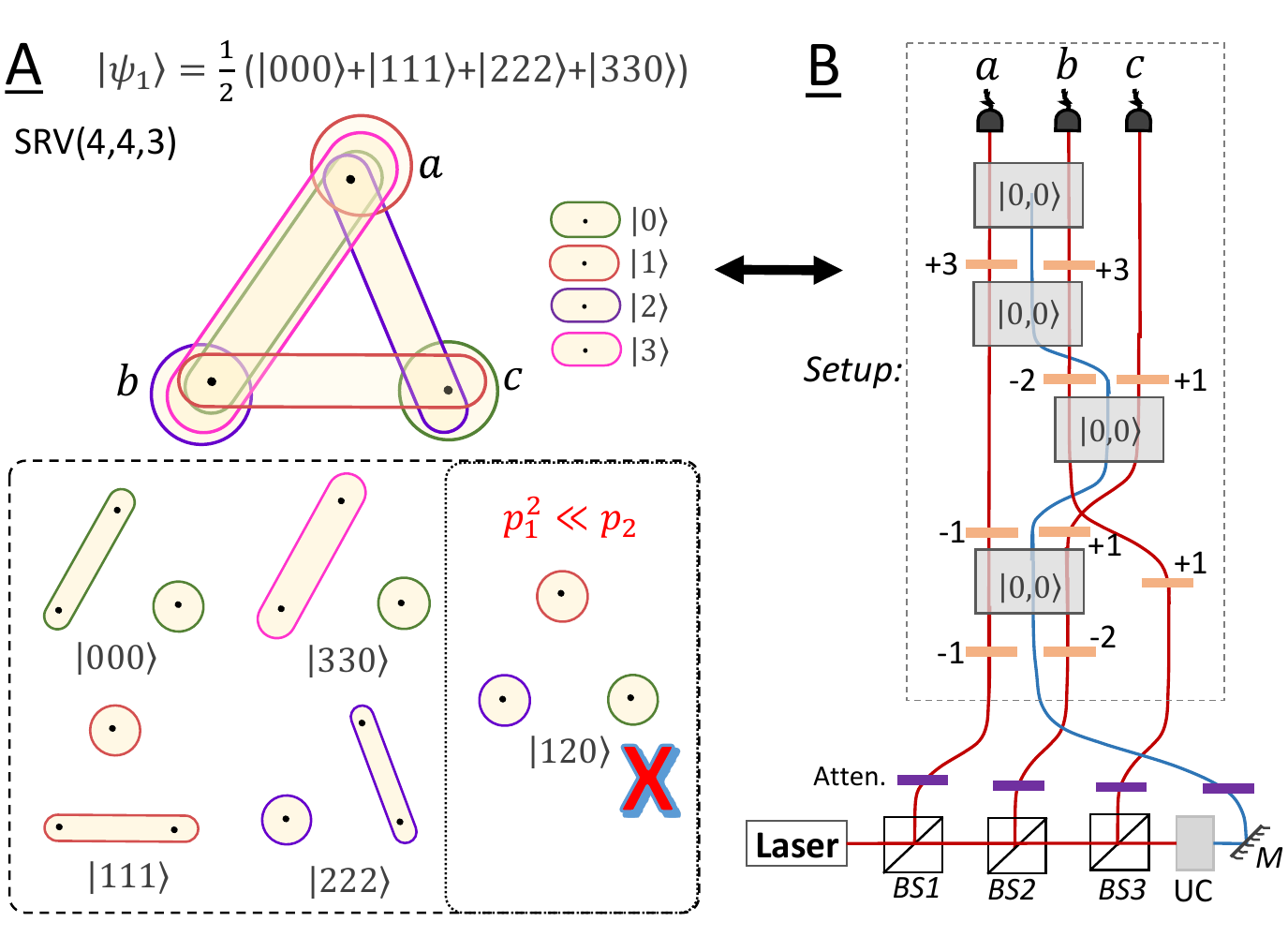}
	\caption{Hypergraph-experiment link for producing state $\ket{\psi_{1}}$. \textbf{A:} Each term of an SRV state $SRV(A,B,C)$ is given by a perfect matching in the corresponding hypergraph. Under the condition of $p_{1}^2 \ll p_{2}$, one can construct a hypergraph of $A$ perfect matchings if $1+min(1+(A-B),C)+min(1+(A-C),B-1)\geq A$ is satisfied \cite{gu2019quantum3}. \textbf{C}: The corresponding setup for creating $SRV(4,4,3)$ state using 1-photon and 2-photon sources.}  
	\label{fig:srvexample}
\end{figure}

As an example, we show a maximally entangled state with $SRV=(4, 4, 3)$, which is
\begin{equation}
\ket{\psi}_{abc}=\frac{1}{2}(\ket{000}+\ket{111}+\ket{222}+\ket{330}).
\label{eq:SRVexample}
\end{equation}
There the first particle $a$ is 4-dimensionally entangled with the other two particles $bc$, particle $b$ is 4-dimensionally entangled with particles $ac$, whereas particle $c$ are only in 3-dimensionally entangled with the rest particles $ab$. Here we are only interested in maximally entangled states, which means all amplitudes are the same. Furthermore, we want that the quantum state with $SRV(A,B,C)$ has A terms. Thereby, the structure of the SRV is clearly visible in the computation basis, which is convenient experimentally. We call such an quantum entangled state an $SRV(A,B,C)$ state. 

The generation of these SRV states has been well-investigated theoretically \cite{krenn2016automated, gu2019quantum3} and experimentally \cite{malik2016multi, erhard2018experimental, hu2020experimental}. In order to make future experimental investigations possible, we aim to find these $SRV(A, B, C)$ state without additional particles with probabilistic $n$-photon sources (in this case $n=1,2$). We exploit our presented hypergraph-experiment connection for experimentally designing and find that the results in \cite{gu2019quantum3} using graph-theoretical concept are applicable using hypergraphs under the restriction $p_{1}^2 \ll p_{2}$. We show one example state described in Eq. \ref{eq:SRVexample}  in Fig. \ref{fig:srvexample}.

Recent advances in integrated optics and silicon photonics have caused increasing attention to entanglement generation on chips \cite{fengprogress, wang2019integrated}. Many chip-based source such as heralded single-photon sources \cite{collins2013integrated, spring2017chip}, high-dimensional states \cite{wang2018multidimensional, lu2019three}, multiphoton states \cite{adcock2019programmable, zhang2019generation, feng2019generation} and on-chip transverse-mode entangled photon pair source \cite{feng2019chip} have been successfully demonstrated. Those platforms are ideal for the practical implementation of our approach.

\subsection{$n$-photon sources: $n\geq 3$}

\begin{figure}[!t]
	\includegraphics[width=0.5\textwidth]{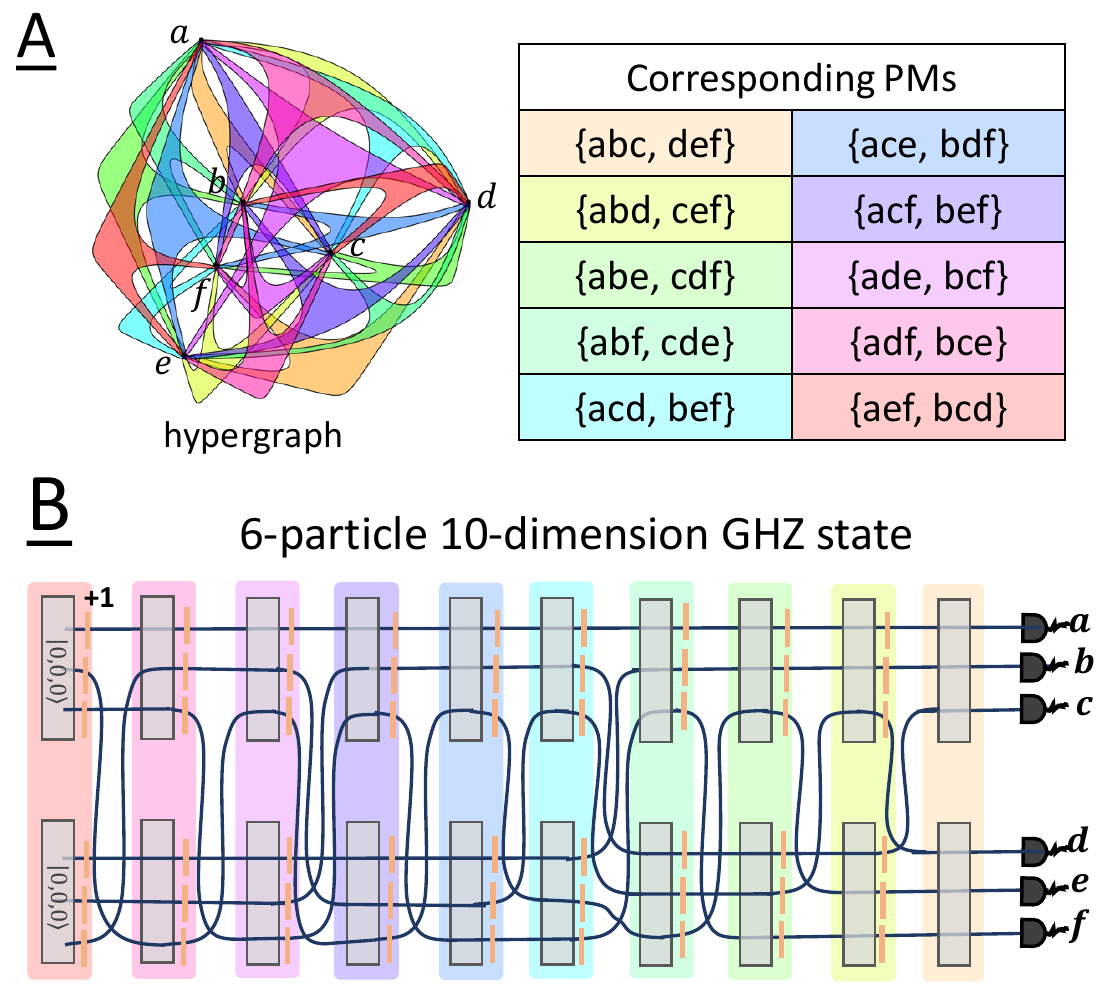}
	\caption{Hypergraph-Experiment description for producing 10-dimension 6-particle GHZ states using 3-photon sources. \textbf{A}: A 3-uniform hypergraph with 20 hyperedges and 6 vertices. There are 10 disjoint perfect matchings (PMs), \textit{every hyperedge only appears in at most one of the perfect matchings}, depicted in the right table with colored backgrounds. \textbf{B}: The corresponding experimental implementation. There hyperedges with $d(e)=3$ and vertices stand for 3-photon sources and output paths, respectively. All sources are pumped coherently and their output paths are aligned identically. The pump power can be set such that two triplet photons are created with reasonable probabilities. The coherent superposition of all perfect matchings gives the post-selected quantum state, which is a 10-dimension 6-particle GHZ state.}  
	\label{fig:Nemitter}
\end{figure}

So far, we have considered the cases of the $n=1$ and $n=2$ photon sources. A general concept of $n$-photon source has been theoretically and experimentally investigated  \cite{shi2015multiphoton, munoz2014emitters, bin2019n}, which can be perfectly interpreted by a hyperedge with $d(e)=n$ in the hypergraph. That means a setup only using $n$-photon sources can be translated into a $k$-uniform hypergraph where $k=n$. For simplicity, we only consider the case $n=3$ and show an example in Fig. \ref{fig:Nemitter}. There are 10 disjoint perfect matchings (i.e., \textit{every hyperedge only appears in at most one of the perfect matchings}) in the corresponding hypergraph, which are depicted in the different colored backgrounds. The final quantum state is created conditioned on $6$-fold coincidences. It can be interpreted as a coherent superposition of perfect matchings that leads to a 10-dimensional 6-photon GHZ state.

If we use 9 output modes with three 3-photon sources firing simultaneously, we can create a 13-dimensional 9-photon GHZ state. Interestingly, this indicates that the dimension of GHZ states grows when more crystals are added in the case of 3-photon sources. This is in stark contrast to the case of 2-photon sources where the maximum dimension $d=3$ \cite{krenn2017quantum, 267013, krenn2019questions}. If we restrict ourselves to two $n$-photon sources firing simultaneously, then the maximal possible dimension for a GHZ state grows as $n=2-7; d=3,10,35,126,462,1716,6435$, which is potentially connected to the integer sequence in \textit{OEIS A001700} \cite{oeisA001700} (number of ways to put $n+1$ indistinguishable balls into $n+1$ distinguishable boxes).

Until now, our hypergraph-experiment connection mainly exploits the recently developed technique -- \textit{entanglement by path identity}  \cite{krenn2017entanglement}, where one cannot determine the origin of every $N$-fold coincidence event. In realistic experimental scenarios, this method requires perfect output path overlapping, suitable temporal coherence and indistinguishability which can be obtained in \cite{jha2008temporal, krenn2017entanglement}. If there are misalignments between the overlapping paths, it does not reduce the coherence between the different terms but changes their relative amplitudes. This is because misaligned beams do not arrive at the detectors and consequently do not lead to a $N$-fold coincidence count. Also,  there are other unavoidable noise such as multiple photon emission, stimulated emission, loss of photons (including detection efficiencies). Similarly as the misalignment case, these noise also reduces the entanglement by unweighting the state. However, those effect could be compensated by adjusting the pump power before each photon source. Therefore maximally entangled, arbitrary, high-dimensional entanglement states should be possibly created experimentally.

Our hypergraph technique can be applied to find experimental implementations for different high-dimensional multipartite quantum states, which will be interesting to study in more detail in future - both for fundamental properties as well as for their applications in novel quantum communication protocols.

\begin{figure*}[!t]
	\includegraphics[width=\textwidth]{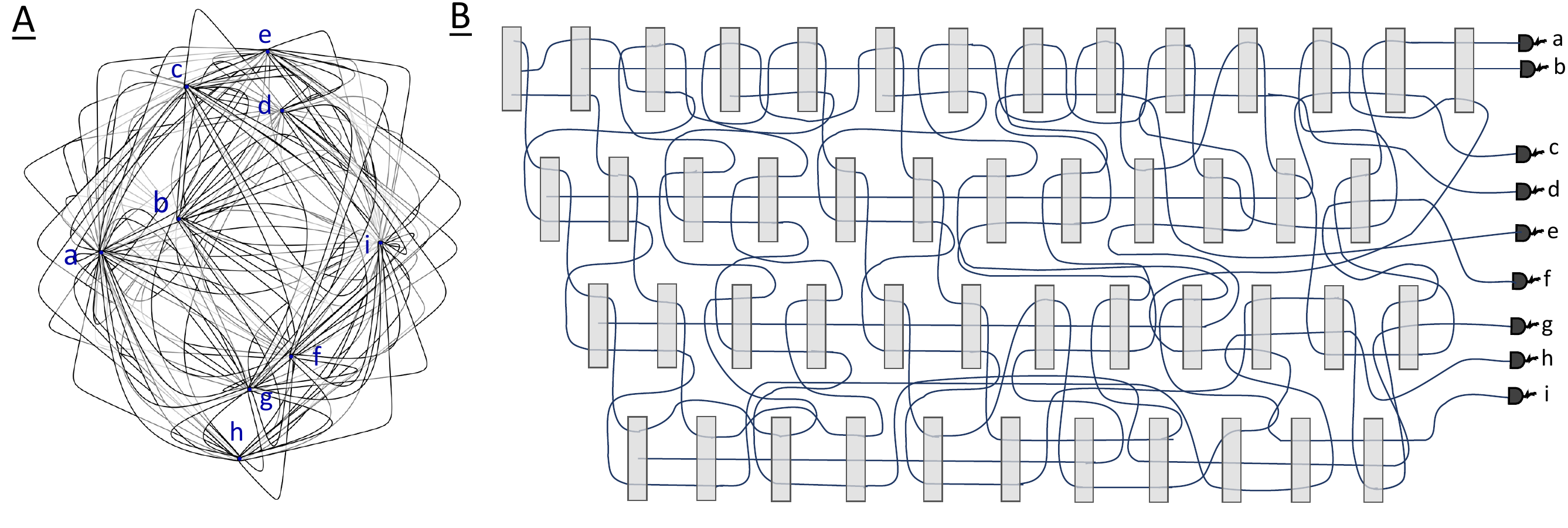}
	\caption{An example showing the difficulty of finding a perfect matching in a 3-uniform hypergraph and its corresponding quantum experiment. \textbf{A}: The 3-uniform hypergraph contains 9 vertices and 49 hyperedges. Interestingly such hypergraph has no perfect matchings while adding more hyperedges can create one perfect matching. \textbf{B}: The 3-uniform hypergraph is transferred into a quantum experiment with multiple 3-photon sources. In this case, there will be no 9-fold coincidence event in the experiment which indicates that there are no perfect matchings in that hypergraph. However, one can just add more 3-photon sources (for example the 3-photon sources' output paths are connecting to $\{a,b,c\}$, $\{d,e,f\}$ and $\{g,h,i\}$ respectively), then there will be a 9-fold coincidences event in the experiment which means there is one perfect matching in the hypergraph by adding more hyperedges. In general, deciding whether a hypergraph has a perfect matching, thus whether a quantum experiment produces $N$-fold coincidences, is \textit{NP-Complete}. This could motivate a hypergraph version of BosonSampling-like quantum surpremacy algorithm, for which a setup similar to the one proposed here could act as the experimental implementation.}  
	\label{fig:HGComplexity}
\end{figure*}

\section{Computation Complexity of Hypergraphs}

In a quantum experiment, the post-selected quantum state is described as the coherent superposition of perfect matchings in its corresponding hypergraph, which means the number of terms in the quantum state corresponds to the number of perfect matchings in the hypergraph. Detecting an $N$-fold coincidences event in a quantum experiment is thus equivalent to guaranteeing a perfect matching in the hypergraph.

The problem of finding a perfect matching in graphs is well understood. For instance, two well-known examples are Tutte's Theorem \cite{tutte1947factorization} and Hall's Marriage Theorem \cite{hall1935j}, which provide necessary and sufficient conditions for the existence of at least one perfect matching in a graph and a bipartite graph respectively. Such graph perfect matching decision problems are efficiently solvable (i.e., Edmonds' algorithm \cite{edmonds1965paths}). These mathematical tools can be employed to find out whether a certain quantum experiment with photon pair sources will produce an $N$-fold coincidence click.

Although the graph perfect matching problem is fairly well-understood, and solvable in polynomial time, most of the problems related to hypergraph perfect matching tend to be very difficult and remain unsolved. Indeed, deciding if a $k$-uniform hypergraph ($k \geq 3$) contains a perfect matching is among the historic 21 \textit{NP-Complete Problems} given by Karp \cite{karp1972reducibility}.

The inability to efficiently decide whether a perfect matching exists means that no efficient classical algorithm can determine from the experimental setup whether there will be an $N$-fold coincidence click at all. This restriction is much stronger than experiments with 2-photon sources. Interestingly, one could take advantage of the mathematical difficulty in quantum experiments. One could build the experiment corresponding to the hypergraph in a laboratory and observe whether there are $N$-fold coincidence clicks. It points toward the possibility of classically intractable problems that can be answered using quantum resources. This difficulty is related to, but conceptually simpler than Boson Sampling \cite{aaronson2011computational, lund2014boson, rahimi2015can, bradler2018gaussian, bradler2018graph, kruse2019detailed, brod2019photonic, zhong2019experimental, wang2019boson}. Boson Sampling exploits the fact that counting all perfect matchings is difficult classically (which is in \textit{$\#$P-complete} complexity class \cite{valiant1979complexity}), and obtains related properties using sampling techniques.

For hypergraphs already, deciding whether a perfect matching exists at all is classically difficult and is \textit{NP-Complete}. Several theoretical studies identified sufficient algorithms for finding perfect matchings in hypergraphs, but the general problem is unsolved mathematically\cite{kuhn2013matchings, khan2013perfect, khan2016perfect}. However, experimentally one could just record whether an $N$-fold coincidence count exists. It might be possible to exploit this classical difficulty for new quantum supremacy and quantum computation protocols, potentially by generalizing the system to a hypergraph-version of BosonSampling \cite{harrow2017quantum, lund2017quantum, arute2019quantum}. This could be done by sampling a subset of outputs from the experiment, which is produced by multiple multi-photon sources. In the standard BosonSampling case, calculating one output requires the evaluation of the matrix function Hafnian (a generalisation of Permanent, which lies in the complexity class \#P). For multi-photon sources, calculating the output lies in a the class \#W[1] \cite{liu2016counting}. The general and systematic experimental designs for such setups motivates further theoretical investigation into this problem.

We show now an example that demonstrates the difficulty of identifying perfect matchings in Fig. \ref{fig:HGComplexity}A. This 3-uniform hypergraph with 9 vertices and 49 hyperedges corresponds to an experimental implementation in Fig. \ref{fig:HGComplexity}B. It is very challenging to find whether there is a perfect matching in such a hypergraph. Interestingly, this hypergraph does not have any perfect matchings but adding any new hyperedges will necessarily create one. 

For realistic experimental situations, one needs to carefully consider the influence of multiple photon emissions, stimulated emission, loss of photons (including detection efficiencies), and amount of photon distinguishabilities in connection with statements of computation complexity. A full investigation of these very interesting questions is beyond the scope of the current paper.

\section{Many-particle Interference}

\begin{figure}[!t]
	\includegraphics[width=0.5\textwidth]{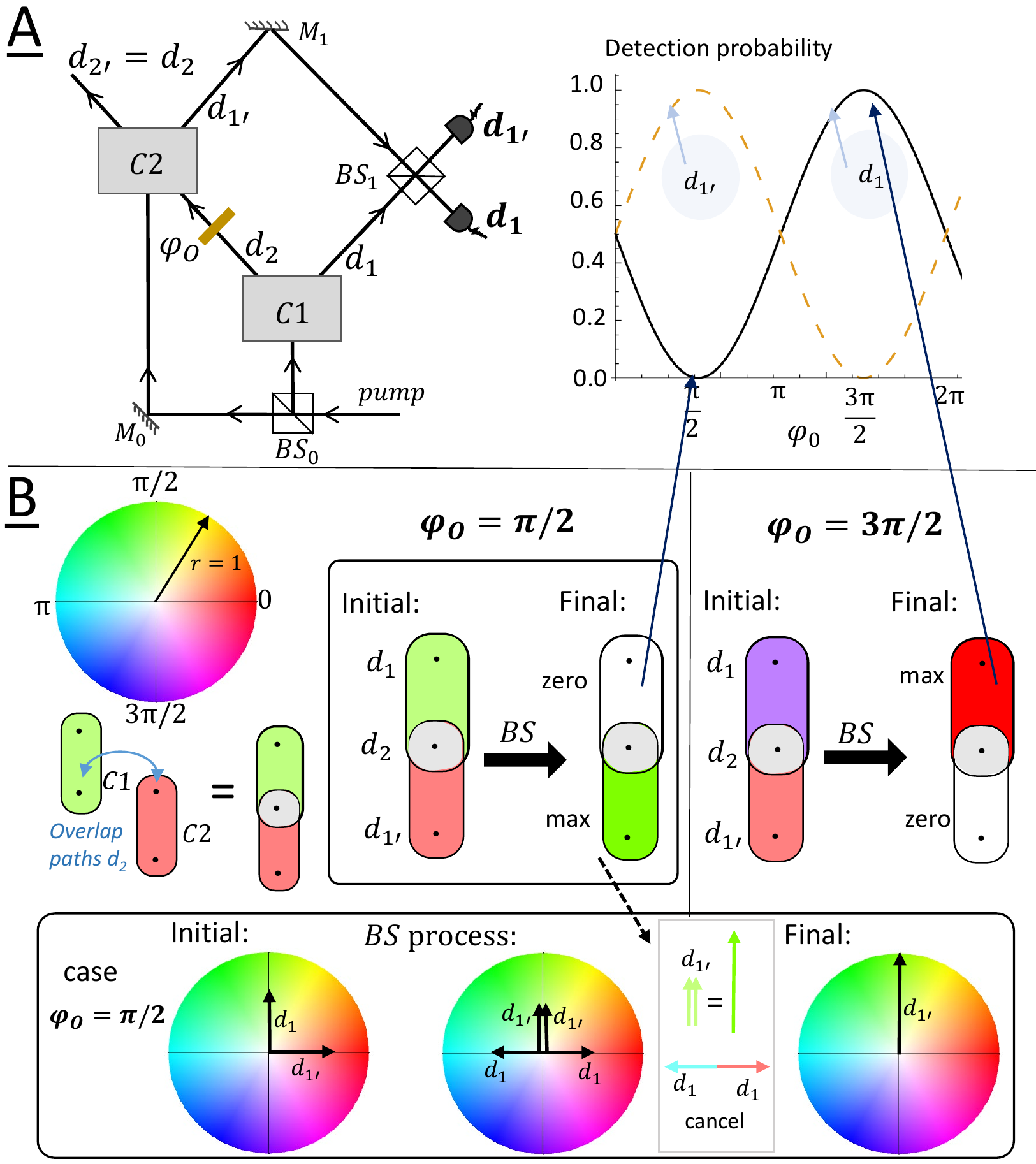}
	\caption{Zou-Wang-Mandel experiment \cite{wang1991induced,zou1991induced} and its hypergraph description. \textbf{A}: Two-photon sources (between them is a phase shifter $\varphi_{0}$) produce photon pairs in paths ($d_{1}$, $d_{2}$) and ($d_{1'}$, $d_{2'}$) respectively and the photon output paths $d_{2}$ and $d_{2'}$ are coherently overlapped. The single-photon interference can be observed when one detects photons in $d_{1}$ or $d_{1'}$ by changing $\varphi_{0}$ without any interaction of photons in path $d_{2}$. \textbf{B}: Hypergraph description for the setup. The complex weights introduced by the phase shifter can be described using a color wheel scale (radius $r$=1). There the phase and amplitude are interpreted by the color and transparency in the region of a hyperedge. We set phases $\varphi_{0}$ to $\pi/2$ and $3\pi/2$ and show the initial and final hypergraphs. In addition, we can directly illustrate the process of the setup in the solid box. The labeled arrow $d_{1}$ in the color wheel stands for the probability of a photon in path $d_{1}$, which is described by the color in the endpoint of the vectors. For better viewpoint, we show it with black arrow for the case of $\varphi_{0}=\pi/2$ (similarly for the case of $\varphi_{0}=3\pi/2$). According to the operation of a BS, rotating the labeled arrow with $\pi/2$ corresponds to reflection. Switching the original label to the label for another input path of BS corresponds to transmission. In the end, the green and white area in the hypergraph are interpreted as labeled arrow in the color wheel, showing the photon in $d_{1'}$ has maximum probability and there is no photon in path $d_{1}$. One can easily see that the probability of detectors $d_{1}$ and $d_{1'}$ are complementary as described in \textbf{A}. That indicates destructive and constructive interference.}
	\label{fig:Mandelgraph}
\end{figure}

A general concept of many-particle interferometry and entanglement based on \textit{path identity} has been presented recently by Lahiri \cite{lahiri2018many}. How can our hypergraph-experiment connection describe such situations? Firstly, we start with the simplest case -- Zou-Wang-Mandel experiment \cite{wang1991induced,zou1991induced} in Fig. \ref{fig:Mandelgraph}A. There two identical 2-photon sources are aligned such that output paths $d_{2}$ and $d_{2'}$ are coherently overlapped. Therefore, one cannot determine the which-source information when observing the photon in path $d_{2'}$. Without interacting photons in path $d_{2'}$, single-photon interference pattern is obtained when photons in $d_{1}$ and $d_{1'}$ are superposed by a 50:50 BS.

Until now, we have familiarized ourselves with the hypergraph-experiment connection for state generations. Now we introduce different complex weights in the hypergraphs, which can naturally describe quantum interference in the experiments. We utilize a color wheel scale of radius $r=1$ to represent a complex weight shown in Fig. \ref{fig:Mandelgraph}B. There we interpret the phase $\varphi$ and transition amplitudes into the color and transparency in the region of a hyperedge. The transparency ratio from 100$\%$ to 0$\%$ describes the related color in radial  from $r=0$ to $r=1$. For example, white color indicates the amplitude is zero (no photon is in the paths) while red color stands for the photon's maximum amplitude and its phase is $0$ or $2\pi$.

We now translate the experiment into its hypergraph description in Fig. \ref{fig:Mandelgraph}B. For simplicity, we only consider two cases that the phase $\varphi$ is set to $\pi/2$ and $3\pi/2$. The actions of linear optics such as \textit{BS-Operation} can be described in graphs \cite{gu2019quantum}, which is extensible in hypergraphs. All linear optical elements such as mode shifter, are feasibly describable in hypergraphs with an internal vertex set. For further details about the graph description of linear optics, see \cite{gu2019quantum}. Here we only show the corresponding initial and final hypergraphs in Fig. \ref{fig:Mandelgraph}B. Clearly, we find that the destructive and constructive interference happens when photons in path $d_{2}$ are never detected. 

\begin{figure}[!t]
	\includegraphics[width=0.5\textwidth]{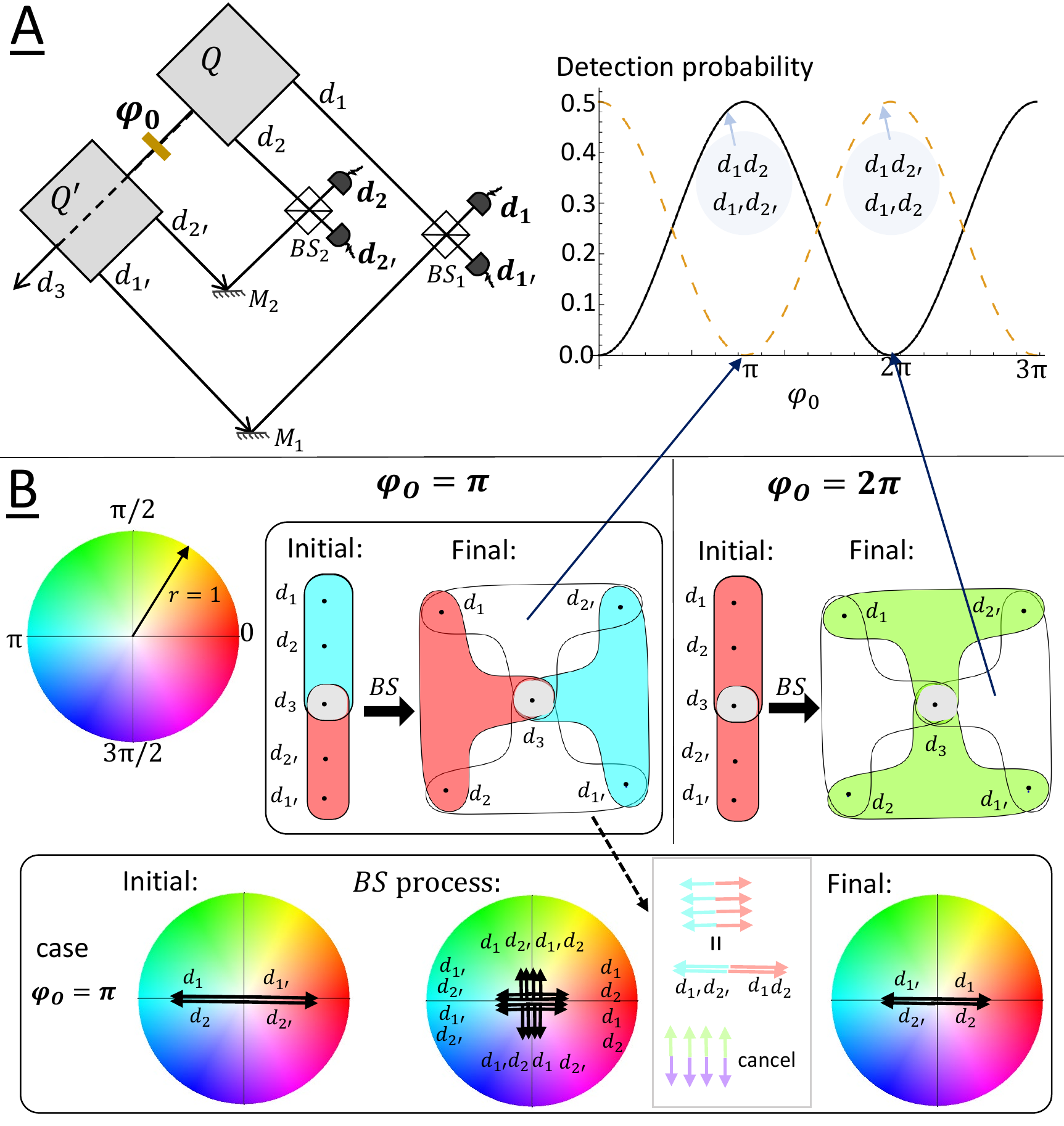}
	\caption{A setup for two-particle interferometry and state generation using 3-photon sources \cite{lahiri2018many} and its hypergraph. \textbf{A}: Two identical 3-photon sources $Q$ and $Q'$ with a phase shifter ($\varphi_{0}$) inserted are pumped coherently. The output paths $d_{3}$ and $d_{3'}$ are aligned identically. Photons in paths ($d_{1}$, $d_{1'}$) and ($d_{2}$, $d_{2'}$) are superposed by two BSs. Without detecting photons in path $d_{3}$, two-particle interference patterns can be observed. Detection probability such as $P_{d_{1}d_{2}}$ denotes the joint detection at pairs of detectors ($d_{1}, d_{2}$). Interference patterns for $P_{d_{1}d_{2'}}$ and $P_{d_{1'}d_{2}}$ vary with phase (dashed line), which are complementary to the patterns for $P_{d_{1}d_{2}}$ and $P_{d_{1'}d_{2'}}$ (solid line). \textbf{B}: Initial and final hypergraphs for the experiment. In analogous to Fig. \ref{fig:Mandelgraph}B, we set the phases $\varphi_{0}$ to $\pi$ and $2\pi$. One can also directly illustrative the setups using the labeled arrows in the color wheel shown in the solid box for the case of $\varphi_{0}=\pi$, similarly for the case of $\varphi_{0}=2\pi$. In the finial hypergraphs, we find that interference patterns $P_{d_{1}d_{2'}}$ and $P_{d_{1'}d_{2}}$ are complementary to the patterns $P_{d_{1}d_{2}}$ and $P_{d_{1'}d_{2'}}$. Furthermore, we find that two different Bell states $\ket{\psi}=\frac{1}{\sqrt{2}}(\ket{d_{1}d_{2}}-\ket{d_{1'}d_{2'}})$ ($\varphi_{0}=\pi$) and $\ket{\psi}=\frac{1}{\sqrt{2}}(\ket{d_{1}d_{2'}}+\ket{d_{1'}d_{2}})$ ($\varphi_{0}=2\pi$) can be obtained when we observe a maximum and a minimum of an interference pattern.}  
	\label{fig:MandelEmittergraph}
\end{figure}

Then we describe the general concept starting with 3-photon sources case. In analogues to Fig. \ref{fig:Mandelgraph}A, two 3-photon sources are pumped coherently and output paths $d_{3}$ and $d_{3'}$ are aligned identically Fig. \ref{fig:MandelEmittergraph}A. Such type of experiment allows one to tune the entangled states and observe many-particle interference patterns without any interaction with the pair of particles. Furthermore, it also provides a new perspective of controlling the amount of entanglement in a quantum state \cite{lahiri2018many}.

Two-particle interference patterns are observed when photons are superposed by two 50:50 BSs. Detection probabilities ($P_{d_{1}d_{2}}$, $P_{d_{1'}d_{2'}}$, $P_{d_{1}d_{2'}}$, $P_{d_{1'}d_{2}}$) at pairs of detectors ($d_{1}, d_{2}$), ($d_{1'}, d_{2'}$), ($d_{1}, d_{2'}$) and ($d_{1'}, d_{2}$) vary sinusoidally with phase $\varphi$. Using our hypergraph-experiment connection, we show the initial and final hypergraphs in Fig. \ref{fig:MandelEmittergraph}B where the phase $\varphi$ is set to $\pi$ and $2\pi$.

Interestingly, the destructive and constructive interference leads to two different Bell states. Specifically speaking, in the case of $\varphi=\pi$ namely destructive interference for ($P_{d_{1}d_{2'}}$, $P_{d_{1'}d_{2}}$), the hypergaph indicates a Bell state $$\ket{\psi}=\frac{1}{\sqrt{2}}(\ket{d_{1}d_{2}}-\ket{d_{1'}d_{2'}}).$$ 
In the case of $\varphi=2\pi$ namely constructive interference for ($P_{d_{1}d_{2'}}$, $P_{d_{1'}d_{2}}$), the hypergraph indicates another Bell state $$\ket{\psi}=\frac{1}{\sqrt{2}}(\ket{d_{1}d_{2'}}+\ket{d_{1'}d_{2}}).$$ Thus a maximum and a minimum of an interference pattern can be attained for two different Bell states. Without interacting with the associated particles, one can modify the entangled states and the interference patterns by changing the phase. Our hypergraph description is well picturesque and visualized for understanding the general concept of many-particle interferometry and entanglement using \textit{path identity} \cite{lahiri2018many, krenn2017entanglement}.
  
\section{Conclusion}

In conclusion, we have introduced hypergraphs, a generalization of graphs, to reinterpret quantum optical experiments using probabilistic $n$-photon sources with the technique \textit{path identity} \cite{krenn2017entanglement} and linear optics. One striking feature is that the concept of hypergraphs is ideal to model different types of correlated photon sources and can therefore be used for designing new quantum experiments. It would be very interesting to understand whether even more complex sources, such as cascaded SPDC down-conversion \cite{hubel2010direct, hamel2014direct} can also be described using our hypergraph-experiment connection.

We also show many experimental implementations to create a vast array of well-defined entangled quantum states (i.e., GHZ states, W states) with the hypergraph-experiment connection. The results indicate the dimensionality of multiphotonic entangled states are going beyond those produced in the experiments only using the 2-photon sources. That advantage sheds light on the producibility of arbitrary quantum states using photonic technology with probabilistic $n$-photon sources.

Moreover, hypergraph states \cite{rossi2013quantum, qu2013encoding} as the generalizations of graph states \cite{hein2004multiparty} have become as powerful resource states for measurement-based quantum computation \cite{raussendorf2003measurement, takeuchi2019quantum}, quantum algorithms \cite{rossi2014hypergraph} and quantum error-correction \cite{balakuntala2017quantum}. Despite the similarity of names, graph states and hypergraph states are not related to the techniques we present. It will be of great interest to establish a connection between hypergraph states and the technology developed here.

Most multiphotonic entangled quantum states are created under the condition of $N$-fold coincidence detection, in which we have focused on the case with one photon per path. That is directly connected to perfect matchings of hypergraphs. Although there are might be multiple photons per path, one can use a photon number filter based on quantum teleportation \cite{wang2015quantum} in each output of the setup to solve. For arbitrary photons per path, it would be a very interesting question for future research exploiting not only perfect matchings, but also more general techniques in matching theory. 
 
We have shown that one can investigate the striking properties of hypergraphs by implementing quantum experiments in laboratories. For example, classically intractable decision problems of perfect matchings in hypergraphs can be solved by experimentally detecting an $N$-fold coincidences case, indicating a potential advantage of quantum experiments and further related to quantum computational supremacy \cite{harrow2017quantum, arute2019quantum, wang2019boson}. Our connection may enable future applications in quantum computation, especially in connection to already existing algorithms employing hypergraphs, e.g., the 3-SAT problem \cite{davis1960computing}. It will be exciting to see the potential of quantum experiments as presented here to solve problems in hypergraph theory that classical computers cannot calculate. 

Finally, we introduce complex weights in the hypergraphs for describing a general concept of many-photon interference using $n$-photon sources together with \textit{path identity} \cite{krenn2017entanglement, lahiri2018many}. The picturesque and visualized approach allows us to observe interference easily and provides us the ability to control entanglement without any interaction with the photons. It would potentially inspire new applications of the hypergraph theory in quantum experiments. Furthermore, description of quantum processes at a more abstract level \cite{coecke2017picturing, abramsky2004categorical} and calculations in quantum optics by representing creation and annihilation operators in a visual way \cite{ataman2018graphical} have been recently studied. A combination of these pictorial approaches with our methods could hopefully improve the abstraction and intuitive understanding of quantum processes.

Our hypergraph-experiment technique works very well with probabilistic $n$-photon sources. In order to escape the restrictions of our method, deterministic quantum sources \cite{michler2000quantum, senellart2017high} would need an adaption of the description, and it is not yet known how to describe active feed forward \cite{ma2012quantum}. Can they be described with hypergraphs? What are the techniques that cannot be described in the way presented here?

\section*{Acknowledgements}
The authors thank Anton Zeilinger, Manuel Erhard, Mayukh Lahiri and Armin Hochrainer for interesting discussions. XG and LC acknowledge support from National Key Research and Development Program of China (2017YFA0303700); The Major Program of National Natural Science Foundation of China (No. 11690030, 11690032); The National Natural Science Foundation of China (No.61771236); China Scholarship Council Scholarship (CSC). M.K. acknowledges support from the Austrian Science Fund (FWF) through the Erwin Schr\"odinger fellowship No. J4309.
 
\bibliographystyle{unsrt}
\bibliography{refs}

\end{document}